# Conservation of the Stokes-Einstein Relation in Supercooled Water

Gan Ren (任淦)[1], and Yanting Wang (王延颋)[2, 3,*]

[1]*Department of Physics, Civil Aviation Flight University of China, Guanghan 628307, China*

[2]*CAS Key Laboratory of Theoretical Physics, Institute of Theoretical Physics, Chinese Academy of Sciences, Beijing 100190, China*

[3]*School of Physical Sciences, University of Chinese Academy of Sciences, Beijing 100049, China*

**ABSTRACT:** The Stokes-Einstein (SE) relation is commonly regarded as being breakdown in supercooled water. However, this conclusion is drawn upon testing the validities of some variants of the SE relation rather than its original form, and it appears conflicting with the fact that supercooled water is in its local equilibrium. In this work, we show by molecular dynamics simulation that both the Einstein and Stokes relations are indeed conserved in supercooled water. The inconsistency between the original SE relation and its variants comes from two facts: (1) the substitutes of the shear viscosity in the SE variants are wavevector-dependent, so it is only a cursory approximation; (2) the effective hydrodynamic radius actually decreases with decreasing temperature, instead of being a constant as assumed in the SE variants. Besides supercooled water, this inconsistency may also exist in other supercooled liquids.

*Introduction.*－The Stokes-Einstein (SE) relation [1] combines the Einstein relation $D = k_B T/\alpha$ and Stokes' formula for hard-sphere particles moving in a viscous fluid $\alpha = C\eta a$, where $D$ is the diffusion coefficient, $k_B$ is the Boltzmann constant, $T$ is the temperature, $\alpha$ is the frictional coefficient, $\eta$ is the shear viscosity, $a$ is the effective hydrodynamic radius, and $C$ is a constant depending on the boundary condition. The SE relation has been successfully applied to many cases, such as colloids [2,3], solutions [4], hard sphere systems [5], and pure liquids [6]. On the other hand, it is also frequently found to be breakdown in complex and supercooled liquids, especially supercooled water [7-11]. However, instead of its original form, the validity of the SE relation in supercooled water was tested by its three variants, $D \sim T/\eta$ [9,10], $D \sim \tau^{-1}$ [12,13], and $D \sim T/\tau$ [7,14], with $\tau$ being the structural relaxation time. The expression $D \sim T/\eta$ becomes a substitute by assuming that $a$ is a constant in homogeneous liquid. Due to the difficulty of accurately determining $\eta$ in molecular dynamics (MD) simulation, $\tau$ is frequently adopted to replace $\eta$, resulting in the variant $D \sim \tau^{-1}$ under a further assumption that $\tau$ has a similar temperature dependence as $\eta/T$ [12,13]. Finally, besides the assumption that $a$ is a constant, the variant $D \sim T/\tau$ comes from the approximate relation $\eta = G_\infty \tau$ [7,14], where $G_\infty$ is the instantaneous shear modulus presumed to be a constant.

Because the above assumptions are *ad hoc*, testing the validity of the SE relation by its variants in supercooled water is questionable. Furthermore, since supercooled water is still in its local equilibrium, it is not unreasonable that the SE relation is actually conserved. In this work, we perform MD simulations with the TIP5P and Jagla water models, respectively, to check the consistency between the original SE relation and its variants in supercooled water. These two water models were adopted because they had previously been used to explore the anomaly and the validity of the SE relation in supercooled water [7,14-16]. Below the simulation results with the TIP5P model are presented in the main text, and those with the Jagla model can be found in Supplemental Material. The simulation results with both models qualitatively agree with each other very well.

*Simulation method.*－All our MD simulations were performed with the GROMACS simulation package [17,18]. The periodic boundary conditions were applied in all three directions of the Cartesian space and the system temperature was kept a constant by the Nosé-Hoover thermostat [19,20]. The particle mesh Ewald algorithm [21] was employed to calculate the long-range electrostatic interactions with a cutoff of 1.2 nm in the real space and the van der Waals interactions were calculated directly with a truncated spherical cutoff of 1.2 nm. The system simulated with the TIP5P model consists of 2048 water molecules with a constant density of $\rho$ = 0.976 g/cm$^3$. Twenty-five simulated temperatures are distributed in the range of 240–390 K. The simulation time step is 1 fs and the simulated MD steps for each case range from $10^6$ to $10^8$ depending on the temperature.

*Breakdown of variants and conservation of the Einstein relation.*－To examine the three variants, we have to first determine the diffusion coefficient $D$, structural relaxation time $\tau$, shear viscosity $\eta$, and frictional coefficient $\alpha$. The diffusion coefficient is calculated via its asymptotic relation with the mean square displacement $D = \lim_{t\to\infty}\langle\Delta\mathbf{r}^2(t)\rangle/6t$, where $\Delta\mathbf{r}(t)$ is particle position displacement and $\langle\rangle$ denotes ensemble average. The calculated $D$ values are plotted in Fig. S1a.

The structural relaxation of water is described by the self-intermediate scattering function $F_s(k,t) = \frac{1}{N}\sum_i^N\langle e^{i\mathbf{k}\cdot\Delta\mathbf{r}(t)}\rangle$, where $N$ is the number of water molecules, wavevector $k$ is usually chosen as where the first maximum of the static structure factor is allocated, which is 24.5 nm$^{-1}$ for the TIP5P model. Since the structural relaxation usually follows an exponential relaxation, $\tau$ is determined by $F_s(k,\tau) = e^{-1}$. The calculated $\tau$ as a function of $T$ with $k$ values within 2.5–24.5 nm$^{-1}$ with an interval of 2 nm$^{-1}$ are plotted in Fig. S2a.

The method proposed by Hess [22] was employed to determine the shear viscosity because of its reliability and fast convergence. In this method, an external force $a_x = A\cos(qz)$ is applied in the $X$ direction, where $A$ is the maximum of $a_x$ and $q = 2\pi/l$ with $l$ being the simulation box size and $z$ being the position in the $Z$ direction. The steady-state solution of the Navier-Stokes equation $\frac{\partial u_x}{\partial t} + u_x\cdot\nabla u_x = -\frac{\eta}{\rho}\nabla^2 u_x + a_x$ is $u_x(z) = V\cos(qz)$, where $V$ is the maximum value of $u_x(z)$. Therefore, the shear viscosity can be calculated by $\eta = A\rho/Vq^2$. Because both $\rho$ and $q$ are constants, we use the ratio $A/V$ to evaluate $\eta$, whose values are plotted in Fig. S3a.

The frictional coefficient $\alpha$ was determined by applying a constant external force $F_e$ on 128 selected molecules and all other particles are treated as the background media. In the linear-response regime, the frictional force on an ion $f_r = \alpha v_r$ is equal to $F_e$ after reaching the non-equilibrium steady state, where $v_r = \lim_{t\to\infty}\langle r(t)\rangle/t$. The frictional coefficient can thus be determined by $\alpha = F_e/v_r$, as plotted in Fig. S4a.

Based on $D$, $\tau$, $\eta$, and $\alpha$ shown in Figs. S1–S4, we fit our simulation data with the three formulas $D\sim\tau^{-\xi_1}$, $D\sim(\tau/T)^{-\xi_2}$ ($\tau$ calculated with $k$ = 24.5 nm$^{-1}$), and $D\sim(\eta/T)^{-\xi_3}$ to test the three variants. Moreover, $D\sim(\alpha/T)^{-\xi_4}$ are fitted to test the Einstein relation. A variant or the Einstein relation is valid if the corresponding exponent $\xi_i \simeq 1.0$ ($i$=1, 2, 3 or 4), and invalid otherwise. To facilitate the determination of the exponents, the logarithm of $D$ versus logarithms of $\tau$, $\tau/T$, $\eta/T$, and $\alpha/T$, respectively, are calculated and plotted in Fig. 1.

It can be seen from Figs. 1a and 1b that both $\xi_1$ and $\xi_2$ deviate significantly from 1.0, indicating breakdown of $D\sim\tau^{-1}$ and $D\sim T/\tau$. Each set of the data can be divided into two parts by a crossover temperature $T_X$, and each part can be well fitted by a fractional form with a different $\xi$ value. The $T_X$ is 285 K for $D\sim\tau^{-\xi_1}$ and 280 K for $D\sim(\tau/T)^{-\xi_2}$, both with $\xi_1, \xi_2 < 1$ at $T < T_X$ and $\xi_1, \xi_2 > 1$ at $T > T_X$. Similar phenomena have been observed by Chen et al [11,13] and Xu et al. [7]. As shown in Fig. 1c, $D\sim(\eta/T)^{-\xi_3}$ also follows two fractional forms with $T_X \approx 280$ K. While $\xi_3 = 0.99$ manifests that $D\sim T/\eta$ is valid at $T > T_X$, $\xi_3 = 0.82$ suggests its breakdown at $T < T_X$. This agrees with not only the experimental observation in water that $\xi_3 = 1$ at high temperatures and $\xi_3 = 0.8$ at low temperatures [10], but also the simulation results with the TIP4P/2005 water model [9]. By contrast, as shown in Fig. 1d, the fitted exponent $\xi_4 = 1.01$ for the whole simulated temperature range is so close to 1.0 that it strongly supports the conservation of the Einstein relation $D = k_B T/\alpha$, naturally consistent with the fact that supercooled water, although metastable, is still in its local equilibrium.

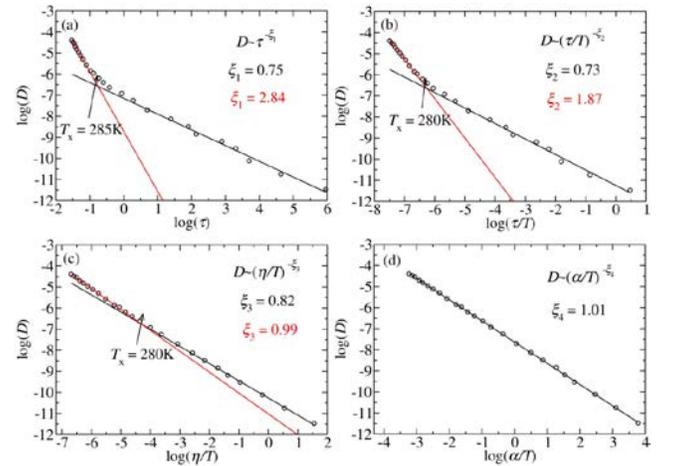

**FIG. 1.** Verification of the validities of the three variants $D\sim\tau^{-1}$ (a), $D\sim T/\tau$ (b), $D\sim T/\eta$ (c), and the Einstein relation $D\sim T/\alpha$ (d). The calculated data are represented by circles and fitted by $D\sim\tau^{-\xi_1}$, $D\sim(\tau/T)^{-\xi_2}$, $D\sim(\eta/T)^{-\xi_3}$, and $D\sim(\alpha/T)^{-\xi_4}$, respectively. The fitted exponent $\xi$ is written in the same color as the corresponding solid fitting line.

*The k dependence of variants.*—Figs. 2a and 2b show the $k$ dependence of $D\sim\tau^{-\xi_1}$ and $D\sim(\tau/T)^{-\xi_2}$, which can be understood as follows. The variant $D\sim\tau^{-1}$ can be exact only if particle displacement follows Gaussian, but it does not even in a system as simple as hard sphere because of the long-time tail effect [23], resulting in the self-intermediate scattering function described by [24]

$$\ln F_s(k,t) = -Dk^2 t + 3(2\gamma/\pi)^{1/2} Dk^2 (t\tau_c)^{1/2} + \cdots \quad (1)$$

where $\gamma = 1$ for pure liquid, $\tau_c = m/\alpha$ is the characteristic time for the Brownian motion, and $m$ is the particle mass. The first term at the right hand side corresponds to the Gaussian case, and the second term corresponds to the memory effect. The contribution of the second term to $\tau$ is usually smaller than the first one, but it does alter $\tau$ from Gaussian, and can be omitted only in the long-wavelength limit $k \to 0$.

The data in Figs. 2a and 2b can also be well fitted by two fractional forms with a crossover temperature, and the corresponding exponents with respect to more $k$ values are shown in Figs. 2c an 2d. Fig. 2c demonstrates that $\xi_{1d}$ decreases while $\xi_{1s}$ increases with decreasing $k$. Both approach 1.0 as $k$ decreases, which agrees with Eq. (1) that $D \sim \tau^{-1}$ is exact only when $k \to 0$. Fig. 2d demonstrates that $\xi_{2s}$ approaches 1.0 as $k \to 0$ at low temperatures, but no such a trend is observed for $\xi_{2d}$ at high temperatures, and thus the basic presumption $\eta = G_\infty \tau$ in $D \sim T/\tau$ is only a cursory approximation. Although it sometimes gives a consistent result with $D \sim T/\alpha$ for a specific $k$ at a certain temperature range, it severely depends on $k$ and does not fulfil for all data.

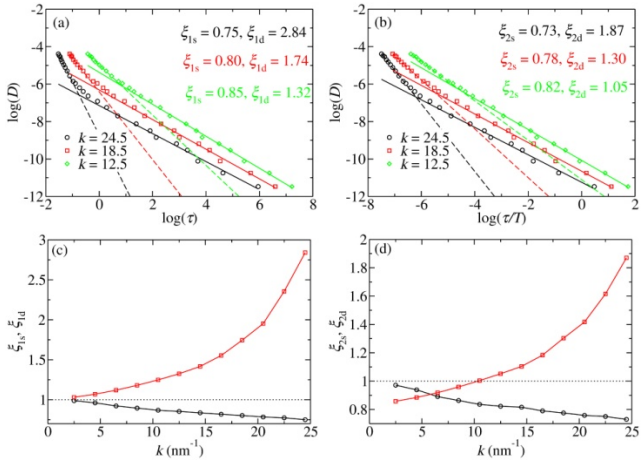

**FIG. 2.** Fitting of the $k$-dependent exponents $\xi_1$ by $D \sim \tau^{-\xi_1}$ (a) and $\xi_2$ by $D \sim (\tau/T)^{-\xi_2}$ (b), as well as $\xi_1$ vs. $k$ (c) and $\xi_2$ vs. $k$ (d). In (a) and (b), the calculated data are represented by different symbols and the fitted exponent is written in the same color as the corresponding fitting line. In (c) and (d), black circles represent $\xi_{1s}, \xi_{1d}$ and red squares are $\xi_{1d}, \xi_{2d}$.

*Variation of $a$ and Conservation of the Stokes relation.*—That $a$ is a constant is the presumption of all three SE variants, but some studies have already shown that it is not a constant [9,10,13]. Below we will show by considering the coordination-shell structure that $a$ in water actually varies with temperature, instead of being a constant.

The coordination shell can be statistically depicted as a water molecule drags the effective shells composed of surrounding water molecules to move together. Here we only consider water molecules in the first coordination shell with an effective radius of $a$, which play the most important role and form a composite along with the central water molecule. The interplay between the central molecule and surrounding water molecules can be described by the coordination number $n$ and the residence correlation time $\tau_s$ [25]. A larger $\tau_s$ corresponds to a larger probability for the molecules to stay in the first shell and move together with the central molecule. The average number of water molecules in the composite is $1 + np(\tau_s)$, where 1 denotes the central molecule and $p$ is the probability of the first shell moving along with the central molecule, which increases monotonically with $\tau_s$. By assuming that both free molecules and the composite are spheres, the effective hydrodynamic radius of the composite is roughly $[1 + np(\tau_s)]^{1/3} a_0$, where $a_0$ is a constant representing the effective hydrodynamic radius of a free molecule. Because the frictional force applied to the composite $\sim C\eta v [1 + np(\tau_s)]^{1/3} a_0$ should be equal to the sum of the frictional force applied to each molecule in the composite $\sim C\eta v [1 + np(\tau_s)] a$, where $v$ is the velocity of the composite, the average effective hydrodynamic radius can be described by

$$a \sim \frac{a_0}{[1 + np(\tau_s)]^{2/3}} \quad (2)$$

With $\tau_s$ described by the Arrhenius law $\tau_s = \tau_0 e^{E_a/k_B T}$, where $E_a$ is the activation energy for a molecule to hop out of the first coordination shell and $\tau_0$ is the prefactor, the probability $p(\tau_s) = \dfrac{e^{E_a/k_B T}}{(e^{E_a/k_B T} + 1)} = \dfrac{\tau_s}{\tau_s + \tau_0}$. Eq. (2) then becomes

$$a \sim \frac{a_0}{\left[1 + \dfrac{n\tau_s}{\tau_s + \tau_0}\right]^{2/3}} \quad (3)$$

Because $\tau_0$ in Eq. (3) is unknown, till now we still cannot determine the temperature dependence of $a$. Therefore, we have to first assume that Stokes' formula is established in supercooled water to numerically evaluate $a \sim \alpha/\eta$, and rescale it by its value at $T = 390$ K for the TIP5P model. The rescaled $\overline{\alpha/\eta}$ plotted in Fig. 3a is approximately a constant for $T > 280$ K but starts to decrease significantly when $T < 280$ K, which has the same trend as $D \sim (\eta/T)^{-\xi_3}$ shown in Fig. 1c and agrees with previous observations [8,9,12]. We then fit the data in Fig. 3a by Eq. (3) with $\tau_0$ being a parameter to be determined. It can be seen that the fitted curve drawn in red in Fig. 3a, which gives $\tau_0 = 269$ ps, almost perfectly matches the numerical data points.

Fig. 3b plots $\tau_s$ and $n$ calculated from the simulation data. When temperature decreases, $n$ decreases because of the more ordered structure of water [7,26], while $\tau_s$ increases because the coupling between molecules becomes stronger. At high temperatures, these two factors contribute oppositely to $np(\tau_s)$ and almost compensate with each other, so the change of $a$ is approximately a constant. At low temperatures, the increase of $\tau_s$ is much faster than the decrease of $n$, so $p(\tau_s)$ plays a more significant role than $n$, leading to the decrease of $a$. The decrease of $a \sim T/D\eta$ with decreasing temperature has been observed in supercooled water by the previous works [9,10,13], and also appears in other liquids, such as supercooled binary Lennard-Jones liquids [27], supercooled aqueous solutions of glycerol [28], water/methanol solutions [29], ortho-terphenyl [30], and tris-Naphthylbenzene [31]. Therefore, we conclude that $a$ indeed varies with thermodynamic condition, which allows the Stokes relation to be valid in supercooled water.

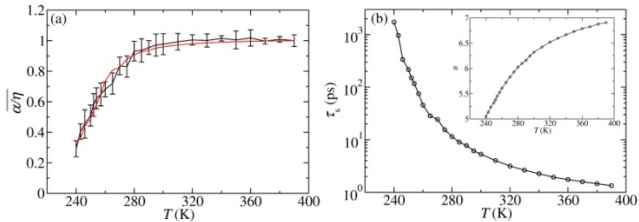

**FIG.3.** (a) Rescaled effective hydrodynamic radius $\overline{\alpha/\eta}$ by simulation (black line) and by fitting (red line). (b) Residence correlation time $\tau_s$ and coordination number $n$ (inset) vs. $T$.

*Conclusion.* ─ We have performed atomistic MD simulations to explore the validity of the SE relation in supercooled water with the TIP5P model in the range of 240–390 K and the Jagla model in the range of 30–140 K. Consistent with the fact that supercooled water is in its local equilibrium, our results confirm the conservation of the original SE relation in supercooled water, even though its three variants, $D \sim \tau^{-1}$, $D \sim T/\tau$, and $D \sim T/\eta$, are all breakdown and in their fractional forms. The variant $D \sim T/\eta$ give rational results only when the effective hydrodynamic radius $a$ can be considered as a constant. In addition, because both $\xi_1$ in $D \sim \tau^{-\xi_1}$ and $\xi_2$ in $D \sim (\tau/T)^{-\xi_2}$ depend on wavevector, $D \sim \tau^{-1}$ agrees with $D \sim T/\alpha$ only in the long-wavelength limit, and $D \sim T/\tau$ agrees with $D \sim T/\alpha$ only in a certain temperature range when a specific $k$ is chosen. Although the three variants give similar qualitative results, the exponents are quantitatively different from each other, so $\tau$ and $\tau T$ are not good substitutes of $\eta$. Overall, the three variants are usually not good substitutes of the original SE relation, and they should be critically and quantitatively evaluated while they are used to inspect the validity of the SE relation. Besides supercooled water, the inconsistency between the original SE relation and its variants may appear in many other supercooled liquids.

The authors thank Prof. Fanlong Meng for his critical reading of this paper. This work was supported by the Strategic Priority Research Program of Chinese Academy of Sciences (Grant No. XDA17010504), the National Natural Science Foundation of China (Nos. 11774357, 12047503) and the Science Foundation of Civil Aviation Flight University of China (Nos. J2021-054, JG2019-19). The allocation of computer time on the HPC cluster of ITP-CAS is also appreciated.

───

* Corresponding author: wangyt@itp.ac.cn.

# Supplemental Material for

## Conservation of the Stokes-Einstein Relation in Supercooled Water

Gan Ren, Yanting Wang*

*Corresponding author. Email: wangyt@itp.ac.cn

### 1. Simulation details for the Jagla model

The initial configuration for the Jagla model [1-3] consists of 1728 molecules with a constant box size determined under a constant pressure $P = 153.33$ bar at each temperature. Although the reduced unit is usually adopted when using the Jagla model, we use the SI unit instead by assigning the unit of hardcore diameter $d = 0.3$ nm, that of potential $U_0 = 100 k_B$, and that of pressure $P_0 = U_0/d^3$. All our molecular dynamics (MD) simulations were performed with the GROMACS MD simulation package [4,5] and the periodic boundary conditions were applied in all three directions of the Cartesian space. The van der Waals interactions were calculated directly with a truncated spherical cutoff of 1.0 nm. Twenty-four temperatures simulated with the Jagla model are in the range of 30–140 K. The system temperature was kept a constant by the Nosé-Hoover thermostat [6,7] and the pressure was controlled by the Parrinello-Rahman barostat [8,9]. The simulation time step is 1 fs and the total simulated MD step for each case ranges from $10^5$ to $10^7$ depending on the temperature. The $k$ values within 2.5–14.5 nm$^{-1}$ with an interval of 1 nm$^{-1}$ for the Jagla model are adopted to investigate the $k$ dependences of $D \sim \tau^{-1}$ and $D \sim T/\tau$. The viscosity $\eta = A\rho/Vq^2$ of Jagla is evaluated by $A/Vl$ because the box size $l$ is different at different temperatures.

### 2. Supplementary Figures

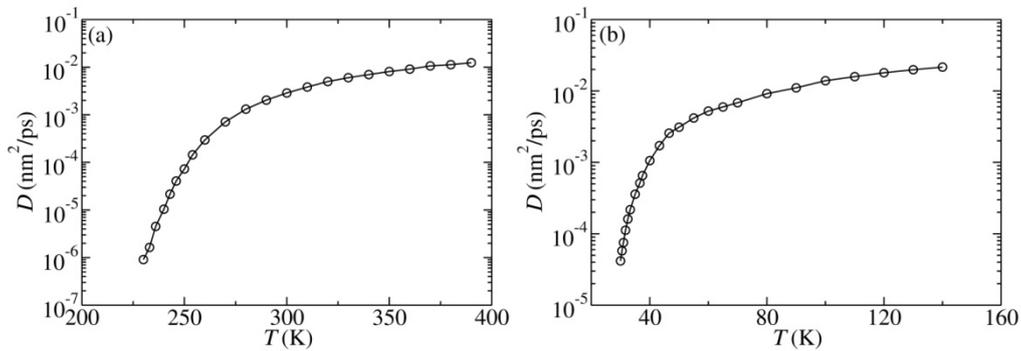

**Fig. S1.** The diffusion constant $D$ as a function of temperature $T$ within 240-390 K for TIP5P (a) and within 30-140 K for Jagla (b).

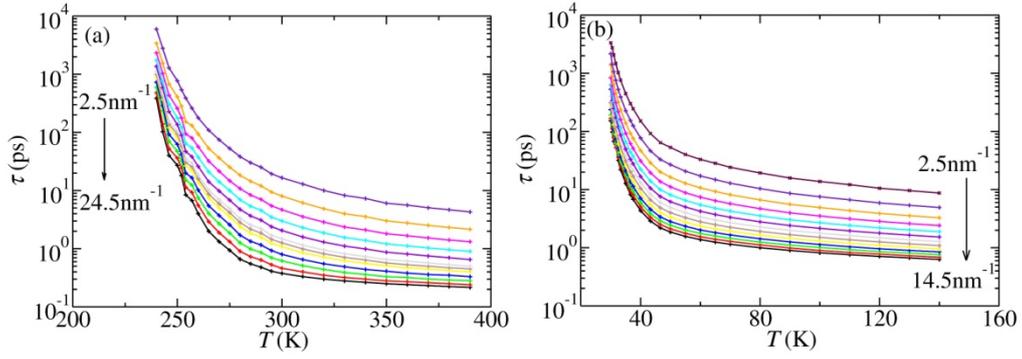

**Fig. S2.** The structural relaxation time $\tau$ for various $k$ values as a function of temperature $T$ for TIP5P (a) and for Jagla (b). The interval between two adjacent $k$ values is 2 nm$^{-1}$ for TIP5P and 1 nm$^{-1}$ for Jagla.

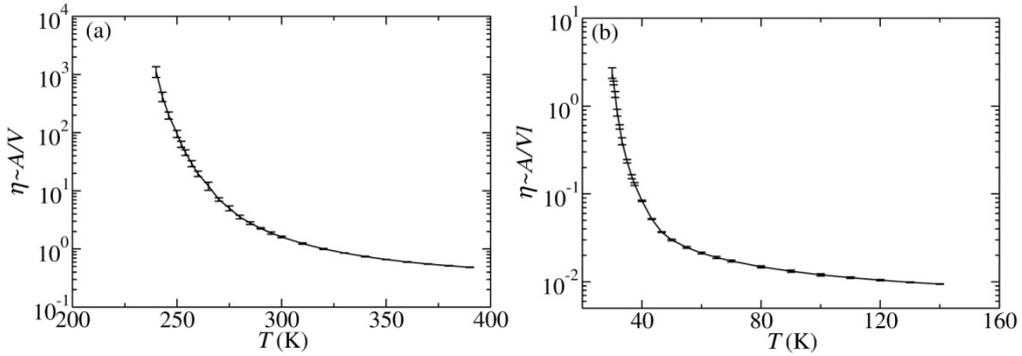

**Fig. S3.** The viscosity $\eta$ as a function of temperature $T$ for TIP5P (a) and Jagla (b).

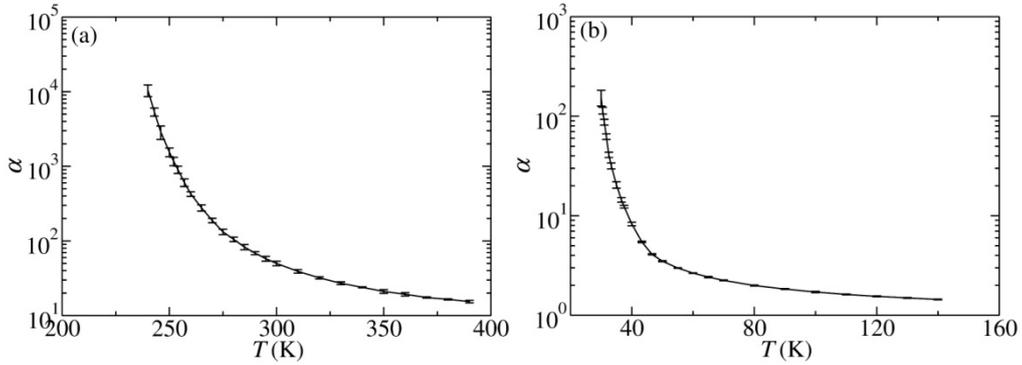

**Fig. S4.** The frictional coefficient $\alpha$ as a function of temperature $T$ for TIP5P (a) and Jagla (b).

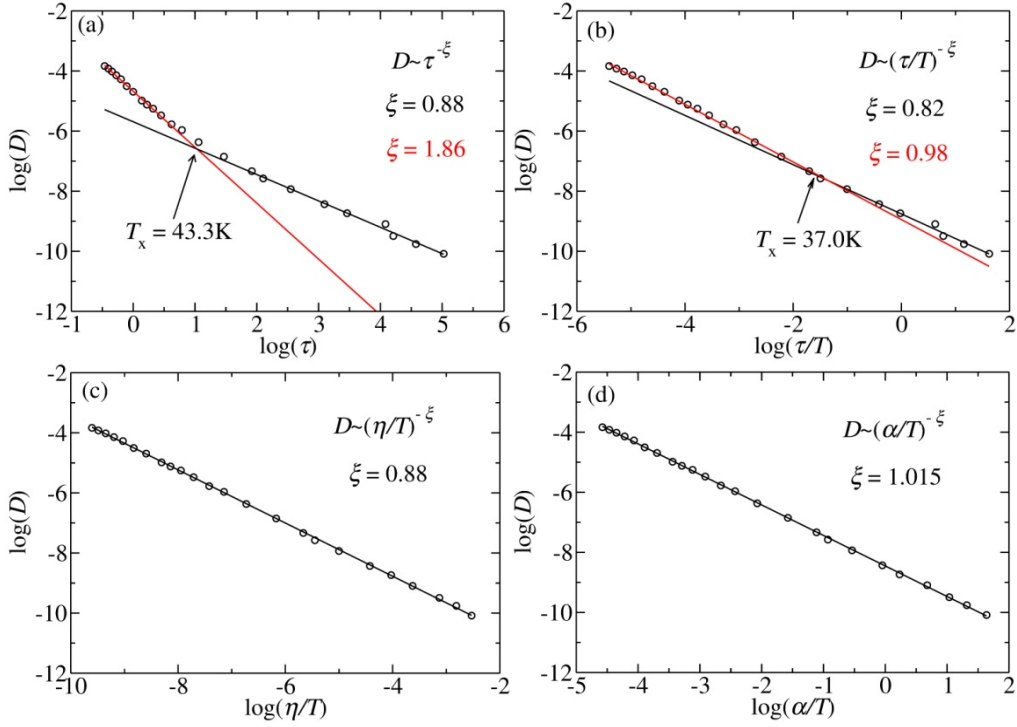

**Fig. S5.** Verification of the validities of the original SE relation $D \sim \tau^{-1}$ (a) and its three variants $D \sim T/\tau$ (b), $D \sim T/\eta$ (c), and $D \sim T/\alpha$ (d) for the Jagla water model. $\tau$ is calculated with the first maximum of static structure factor $k = 12.5$ nm$^{-1}$. The calculated data are represented by circles and fitted by $D \sim \tau^{-\xi_1}$, $D \sim (\tau/T)^{-\xi_2}$, $D \sim (\eta/T)^{-\xi_3}$, and $D \sim (\alpha/T)^{-\xi_4}$, respectively. The fitted exponent $\xi$ is written in the same color as the corresponding solid fitting line.

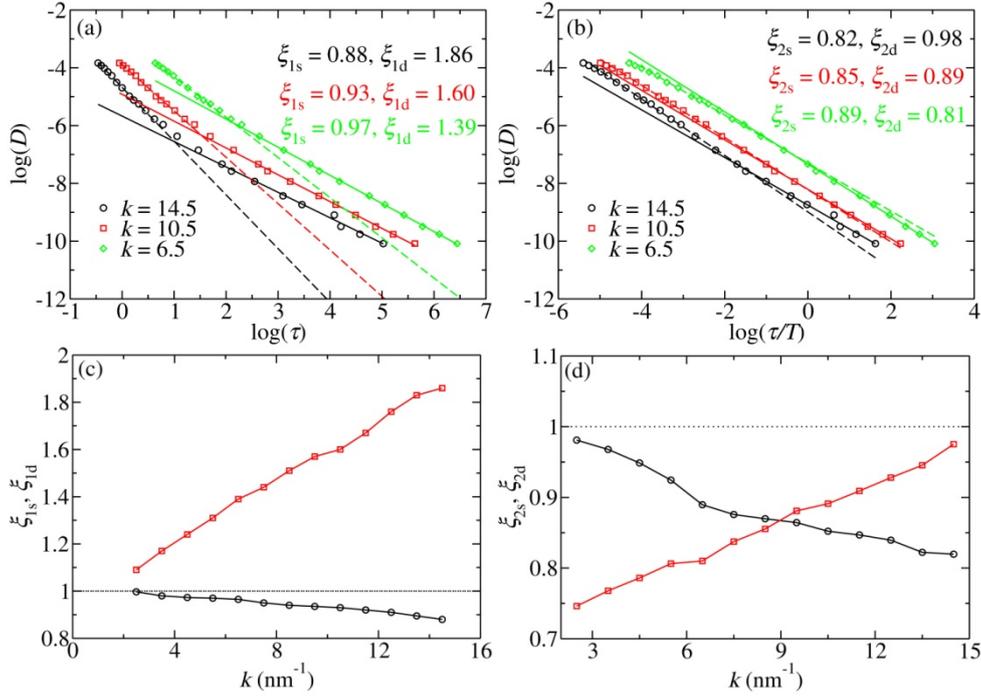

**Fig. S6.** Fitting of the $k$-dependent exponent $\xi_1$ by $D \sim \tau^{-\xi_1}$ (a) and $\xi_2$ by $D \sim (\tau/T)^{-\xi_2}$ for Jagla (b), as well as $\xi_1$ vs. $k$ (c) and $\xi_2$ vs. $k$ (d). In (a) and (b), the calculated data are represented by different symbols and the fitted exponent is written in the same color as the corresponding fitting line; $\xi_{1s}, \xi_{2s}$ values correspond to solid lines and $\xi_{1d}, \xi_{2d}$ values correspond to dotted lines. In (c) and (d), black circles represent $\xi_{1s}$, $\xi_{2s}$ and red squares are $\xi_{1d}$, $\xi_{2d}$.

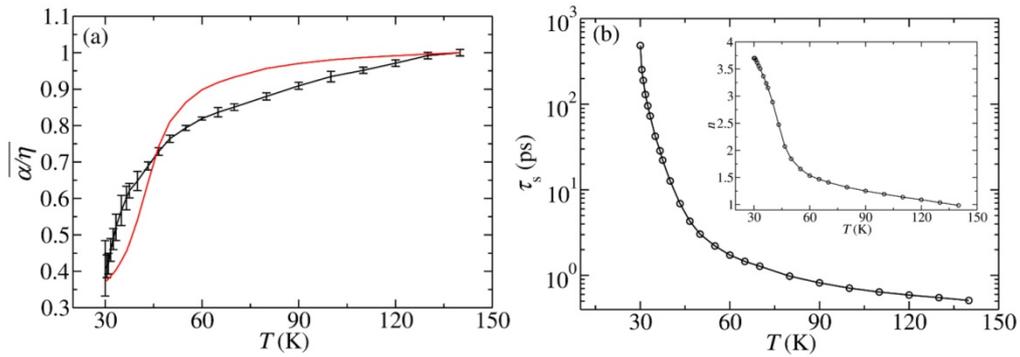

**Fig. S7.** (a) Effective hydrodynamic radius rescaled by the value at $T = 140$ K $\overline{\alpha/\eta}$ vs. $T$ for the Jagla model. The black line is the simulated result and the red one is the fitted result with $\tau_0 = 19$ ps. (b) Residence correlation time $\tau_s$ and coordination number $n$ (inset) of the first solvation shell vs. $T$ for the Jagla model.